\documentclass[aps,pre,
reprint,floatfix,amsmath,amssymb,longbibliography]{revtex4-2}
\usepackage{graphicx}
\usepackage{hyperref}
\hypersetup{hidelinks}
\newcommand{\artanh}{\operatorname{artanh}}
\newcommand{\etad}{\eta_{\mathrm{disk}}}

\begin{document}
\title{Inferring Entropy Production from Correlation Asymmetry}
\author{Haobo Ma}
\affiliation{ChronoAI Pte Ltd, Singapore}
\author{Wenlin Zhang}
\affiliation{National University of Singapore, Singapore 119077, Singapore}
\affiliation{ChronoAI Pte Ltd, Singapore}
\author{Jie Gu}
\email{jiegu1989@gmail.com}
\affiliation{Chengdu Academy of Educational Sciences, Chengdu 610036, China}
\begin{abstract}
Finite-state observables can conceal thermodynamic currents even when their stationary correlations reveal a preferred direction of fluctuation.
We derive a sharp nonlinear lower bound on stationary entropy production using short-time correlation asymmetry, combined autocorrelation decay, and the count and enclosing geometry of observable values.
A uniformly driven ring saturates the bound.
For oscillatory relaxation modes, the result yields a coherence-dissipation constraint that retains information about the distribution of mode amplitudes.
Our results strengthen earlier disk-radius and mode-weighted estimates.
An extension to arbitrary stationary time lags uses endpoint correlations.
\end{abstract}
\maketitle

\section{Introduction}\label{sec:introduction}

Entropy production quantifies the irreversibility of stochastic dynamics. For a stationary Markov jump process, it is determined by the imbalance between forward and reverse transition fluxes \cite{Schnakenberg1976,LebowitzSpohn1999,Seifert2012,Seifert2019}. Direct evaluation therefore requires dynamical information that is rarely available in a partially observed experiment. A biochemical network or mesoscopic device may have many microscopic states, while the measurement records only a few fluorescence, conformational, or other state-dependent signals. Different microscopic states can produce the same reading, and the resulting coarse graining can hide substantial dissipation \cite{Esposito2012,Bisker2017,Martinez2019,SkinnerDunkel2021}. The operational question is then not how to reconstruct every hidden transition, but how much entropy production is necessarily implied by the statistics that can actually be measured. For broader reviews and perspectives on thermodynamic inference and coarse-grained irreversibility, see Refs.~\cite{Seifert2019,Seifert2026Review,GhosalBisker2026,DieballGodec2025Perspective,Alemany2015,GodecMakarov2023}.

Several approaches have turned partial observations into rigorous thermodynamic estimates. Thermodynamic uncertainty relations bound entropy production through the precision of integrated currents \cite{BaratoSeifert2015,Gingrich2016,Pietzonka2016,HorowitzGingrich2017,Dechant2018}. Extensions include optimized short-time estimators and bounds for more general counting observables \cite{Otsubo2020,Shiraishi2021,PietzonkaCoghi2024}. Another line of work recovers irreversibility from coarse-grained paths, visible transitions, waiting-time statistics, and trajectory segments \cite{VanderMeerErtelSeifert2022,HarunariDutta2022,VanderMeerDeguntherSeifert2023,Maier2024,MaierSeifertVanderMeer2025,GuExcursion2026}. These methods use different experimental information and can be powerful when sufficiently resolved trajectories or current-like observables are available. In many measurements, however, it is easier to estimate temporal correlations of a small set of state variables than to identify individual transitions. This motivates bounds expressed directly in terms of correlations rather than reconstructed currents or paths.

For two state observables, equilibrium detailed balance implies symmetry of stationary cross-correlations upon exchanging the observables. The corresponding time-order asymmetry therefore provides a measurable signature of irreversible circulation. Ohga, Ito, and Kolchinsky bounded this asymmetry through cycle affinities \cite{OhgaItoKolchinsky2023}, while Shiraishi related its short-time growth, normalized by autocorrelation decay, to entropy production and the geometric extent of the observable values \cite{Shiraishi2023}. Gu obtained complementary activity- and dissipation-dependent bounds \cite{Gu2024}. Related work has constrained finite-lag asymmetries and coarse-grained time reversal \cite{LiangPigolotti2023,VanVuVoSaito2024}, as well as correlation times and integrated covariances \cite{DechantGarnierBrunSasa2023,AslyamovEsposito2026}. Most recently, Raux, Maier, and Seifert developed entropy-production bounds from correlation asymmetry at arbitrary stationary lag, using additional correlations of observable products, with a separate short-lag result for time-dependent dynamics \cite{RauxMaierSeifert2026}. Collectively, these results establish correlations as an experimentally useful source of thermodynamic information, while also showing that the available bounds depend crucially on which features of the observed signals are retained.

One such feature has not been fully exploited in the short-time inference problem: the observables may take only finitely many distinct {joint} values. This number reflects the resolution of the measurement, not necessarily the number of microscopic states. Together with the radius of the smallest disk containing those values, it restricts the geometry of the closed paths traced by stationary probability currents in the plane of the observables. A disk-radius bound by itself does not use the number of allowed readouts \cite{Shiraishi2023}. This leads to two precise questions. {What is the minimum stationary entropy production compatible with a measured initial cross-correlation asymmetry and combined autocorrelation decay, when the number of distinct joint values and their enclosing-disk radius are specified? Can the resulting lower bound be attained for every admissible set of these data, or does the inference necessarily leave a gap?} The second question is essential: a valid inequality need not be the strongest one supported by the stated information.

We answer both questions for finite, irreducible Markov jump processes with bidirectional allowed transitions. Combining the geometric restriction imposed by finite observable support with the exact nonlinear relation between current and thermodynamic affinity, we derive an explicit lower bound and its admissible domain. The bound is {sharp for the specified information set}: a uniformly driven ring with the joint observable values arranged on a regular polygon attains equality for every admissible nonzero asymmetry. Thus no uniformly stronger bound can be inferred from these summary measurements alone. For finite support, the result improves the disk-radius inequality of Shiraishi through both a support-dependent geometric factor and a nonlinear correction \cite{Shiraishi2023}. 

%Two extensions illustrate the scope of this inference principle. First, retaining only the initial and final states of an observation interval yields a bound at arbitrary stationary time lag. It uses the same finite-support information and reduces to the sharp result in the short-time limit. At nonzero lag, however, discarded intermediate trajectories generally prevent equality. Second, choosing the real and imaginary parts of an oscillatory eigenfunction as observables gives a coherence--dissipation constraint weighted by the mode's stationary amplitude distribution. Earlier studies have related oscillatory coherence to driving in Markov spectra \cite{BaratoSeifert2017,OberreiterSeifertBarato2022,KolchinskyOhgaIto2024,Shiraishi2026} and in weak-noise limit cycles \cite{RemleinWeissmannSeifert2022,SantolinFalasco2025}, under different definitions and assumptions. Our bound retains the eigenfunction information emphasized in the rigorous mode-weighted result of Ref. \cite{Gu2026}.

\section{Results}\label{sec:results}
\subsection{Setup}\label{subsec:setup}

Consider an irreducible continuous-time Markov jump process on $N$ states with the transition rate matrix $R$, where $R_{ij}$ is the rate for a jump {from $j$ to $i$}:
\begin{equation}\label{eq:transition-rate-matrix}
\frac{dp_i(t)}{dt}=\sum_jR_{ij}p_j(t),\qquad
\sum_iR_{ij}=0.
\end{equation}
Off-diagonal rates are nonnegative, and the diagonal entries are fixed by probability conservation.  The steady-state probabilities $p_i^{\rm ss}>0$ obey $\sum_jR_{ij}p_j^{\rm ss}=0$ and $\sum_i p_i^{\rm ss}=1$.
We assume that each allowed jump has a reverse jump: $R_{ij}>0$ if and only if $R_{ji}>0$.
For an edge with the reference direction $i\to j$, define its two steady-state fluxes, net probability current, and dynamical activity (traffic) as
\begin{equation}\label{eq:fluxes}\begin{aligned}
F_{ij}&=R_{ji}p_i^{\rm ss},\qquad F_{ji}=R_{ij}p_j^{\rm ss},\\
J_{ji}^{\rm ss}&=F_{ij}-F_{ji},\qquad
A_{ij}=F_{ij}+F_{ji}.
\end{aligned}\end{equation}
Thus $J_{ji}^{\rm ss}$ is positive for net flow from $i$ to $j$, while $A_{ij}$ measures the overall frequency of jumps along the edge, independent of direction.
In units of Boltzmann's constant, the steady-state entropy production rate is
\begin{equation}\label{eq:sigma}
\dot{\sigma}=\sum_{i\to j}J_{ji}^{\rm ss}\ln(F_{ij}/F_{ji}).
\end{equation}
Each edge contributes a nonnegative amount.  Under local detailed balance this is the physical rate of entropy production \cite{Schnakenberg1976,LebowitzSpohn1999}.

Let $a_i$ and $b_i$ be the values of two state observables.  As in Ref. \cite{Shiraishi2023}, we characterize their short-time fluctuations by the decay rates of their autocorrelations $D_a$ and $D_b$, and by the stationary fluctuation oscillation $\alpha_{ab}$:
\begin{equation}\label{eq:DW-def}\begin{aligned}
D_a&=-\langle a\dot a\rangle
=\frac12\sum_{i\to j}A_{ij}(a_j-a_i)^2,\\
D_b&=-\langle b\dot b\rangle
=\frac12\sum_{i\to j}A_{ij}(b_j-b_i)^2,\\
\alpha_{ab}&=\frac12\langle a\dot b-b\dot a\rangle\\
&=\frac12\sum_{i\to j}J_{ji}^{\rm ss}
(a_i b_j-b_i a_j).
\end{aligned}\end{equation}
Here a dot denotes the initial time derivative of a stationary two-time correlation, not a differentiable sample trajectory.  The autocorrelation decay rates measure how strongly jumps change the observed variables.  The fluctuation oscillation $\alpha_{ab}$ instead measures their directional rotation in the $a$--$b$ plane and vanishes under detailed balance.
For later use, combine the observables as $u_i=a_i+\mathrm{i}b_i$ and write $D(u)\equiv D_a+D_b$ and $W(u)\equiv 2\alpha_{ab}$.  Their steady-state expressions are
$D(u)=\frac12\sum_{i\to j}A_{ij}|u_j-u_i|^2$ and
$W(u)=\sum_{i\to j}J_{ji}^{\rm ss}\operatorname{Im}(\overline u_i u_j)$.
We simply write $D$ and $W$ when referring to the measured pair.
We denote by $d$ the number of distinct values taken
by $u_i$ across all states, which counts different reported values, not necessarily different microscopic states.
It may be less than $N$ if multiple states produce the same pair of readings.

With $C_{fg}(t)=\langle f(X_0)g(X_t)\rangle_{\rm ss}$, the experimentally accessible initial slopes are
\begin{equation}\label{eq:correlation-slopes}\begin{aligned}
D_a&=-C'_{aa}(0^+),\qquad D_b=-C'_{bb}(0^+),\\
2\alpha_{ab}&=C'_{ab}(0^+)-C'_{ba}(0^+).
\end{aligned}\end{equation}
Appendix~\ref{app:slopes} derives these identities directly from the steady-state master equation.
If $D_a+D_b=0$, both observables are constant on every allowed transition, and therefore $\alpha_{ab}=0$.

Only the distinct jointly attainable pairs $(a_i,b_i)$ enter the support geometry.
Define
\begin{equation}\label{eq:support-geometry}
 \rho=\min_{c\in\mathbb C}\max_i|u_i-c|.
\end{equation}
Thus $\rho$ is the radius of their smallest enclosing disk in the complex plane.
Shifting all $u_i$ by the same complex constant leaves $D$ unchanged and, by conservation of stationary current, also leaves $W$ unchanged.
Under a common nonzero rescaling $u\mapsto\lambda u$, both $D$ and $W$ scale by $|\lambda|^2$, as does $\rho^2$.

\subsection{Sharp correlation-based bound from finite observable support}\label{subsec:observable}

\subsubsection{Main bound}

\begin{figure}[t]
\centering
\includegraphics[width=\columnwidth,height=0.38\textheight,keepaspectratio]{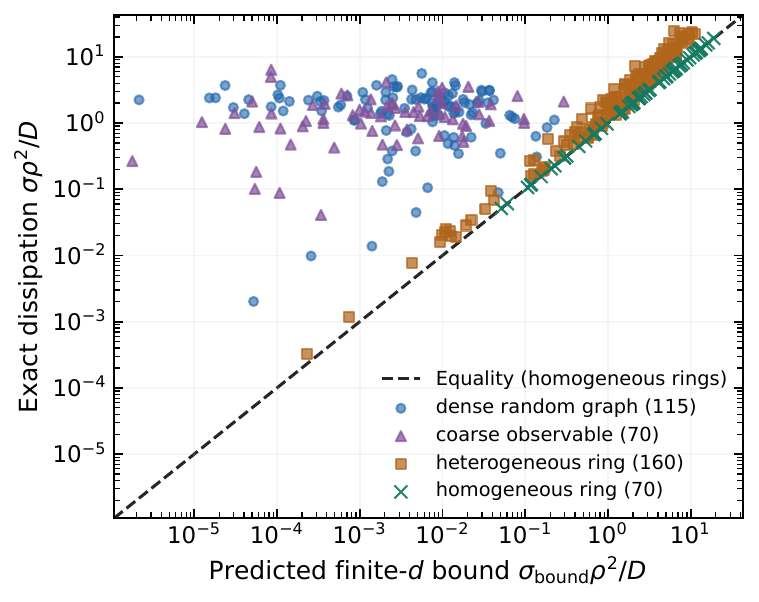}
\caption{Numerical validation of the finite-support EPR bound.
The horizontal axis is the predicted bound and the vertical axis is the
exact EPR, both multiplied by $\rho^2/(D_a+D_b)$. The dashed line is
equality. Both axes are logarithmic. The 415 networks (seed $20261021$)
include 70 homogeneous rings (crosses), 160 heterogeneous rings (squares),
115 complete graphs with distinct levels (circles), and 70 complete graphs
with repeated levels (triangles).
The number of states $N$ is uniform on $\{3,\ldots,11\}$, except for the
repeated-level graphs where $N\geq4$.
For rings, $k_+=\exp U(-0.9,1.1)$ and
$k_-=k_+\exp U(-3.5,-0.12)$.
Homogeneous rings have observable values at the $N$th roots of unity.
In heterogeneous rings each oriented rate is independently multiplied
by $\exp\mathcal N(0,0.36^2)$, and the observable radii and phases are
$\exp\mathcal N(0,0.17^2)$ and
$2\pi i/N+\mathcal N(0,0.13^2)$.
For complete graphs, every directed rate follows $\exp U(-2.8,2.1)$.
Observable radii and phases follow $\exp U(-1.2,0.65)$ and
$U(-\pi,\pi)$. For repeated-level networks, $d$ is uniform on
$\{3,\ldots,N-1\}$.}
\label{fig:scatter}
\end{figure}

When $\alpha_{ab}\ne0$, there must be at least three distinct observed values because a two-vertex closed walk cannot enclose an oriented area.
Moreover, stationary current conservation allows the probability currents to be decomposed into directed cycles. Combining the geometric constraint on the oriented area of each cycle with the fact that the net current is smaller than the dynamical activity on every edge, we obtain
\begin{equation}\label{eq:observable-domain}
   \frac{2|\alpha_{ab}|}{D_a+D_b} <\cot(\pi/d).
\end{equation}
For the whole range, the stationary entropy production rate obeys
\begin{equation}\label{eq:observable-main}
{\dot{\sigma}\geq \frac{4|\alpha_{ab}|}{\rho^2\sin(2\pi/d)}\artanh\!\left(\frac{2|\alpha_{ab}|}{D_a+D_b}\tan\frac{\pi}{d}\right).}
\end{equation}
Here $\artanh z=\tfrac12\ln[(1+z)/(1-z)]$ for $|z|<1$.
This restriction also ensures that the inverse hyperbolic tangent remains finite and real.
Detailed derivations of Eqs.~\eqref{eq:observable-domain} and \eqref{eq:observable-main} are given in Appendix~\ref{app:observable}.

The different measurements play complementary roles.
The ratio $2|\alpha_{ab}|/(D_a+D_b)$ compares directional fluctuation oscillations with the loss of autocorrelation, the finite number $d$ limits how much oriented area a cycle of observable values can enclose, and $\rho$ fixes the geometric scale.
The nonlinear factor captures how a stronger steady-state current requires a larger thermodynamic driving force.
%Equation~\eqref{eq:observable-main} is attained at fixed $D_a+D_b$, $\alpha_{ab}$, $\rho$, and $d$: a homogeneous $d$-state ring with the observable values at the vertices of a regular polygon attains equality for every admissible $q$.

\subsubsection{Saturation}

We now give a uniform ring whose entropy production exactly reaches the lower bound in Eq.~\eqref{eq:observable-main}.
Fix any admissible data $D>0$, $\rho>0$, integer $d\geq3$, and $W\ne0$ satisfying $|W|/D<\cot(\pi/d)$.
Write
\begin{equation}\label{eq:ring-parameters}
\theta=\frac{2\pi}{d},\qquad
h=1-\cos\theta,\qquad
b=\frac{|W|}{D}\tan\frac\theta2.
\end{equation}
The strict domain ensures $0<b<1$.
Take a nearest-neighbor $d$-state ring with periodic indices and choose the observable values and clockwise/anticlockwise rates to be
\begin{equation}\label{eq:ring-construction}
u_k=\rho e^{\mathrm{i}k\theta},\qquad
k_\pm=\frac{D}{2\rho^2h}(1\pm b).
\end{equation}
Both rates are strictly positive, so the bidirectionality assumption is satisfied.
Translational symmetry implies the unique stationary distribution is uniform, $p_k^{\rm ss}=1/d$.
The $d$ observable values form a regular polygon whose smallest enclosing disk has radius $\rho$.
Every edge connects adjacent vertices, so
$|u_{k+1}-u_k|^2=2\rho^2h$ and
$\operatorname{Im}(\overline u_k u_{k+1})=\rho^2\sin\theta$.
The steady-state edge current and dynamical activity are $J=(k_+-k_-)/d$ and $A=(k_++k_-)/d$.
Substitution into the definitions of $D$ and $W$ therefore gives
\begin{equation}\label{eq:ring-DW-check}\begin{aligned}
D(u)&=\rho^2h(k_++k_-)=D,\\
|W(u)|&=\rho^2(k_+-k_-)\sin\theta\\
&=D b\frac{\sin\theta}{h}=|W|.
\end{aligned}\end{equation}
In the last equality we used
$\sin\theta/h=\cot(\theta/2)$ and the definition of $b$.
If the prescribed $W$ is negative, replace the polygon values $u_k$ by their complex conjugates (equivalently, reverse the polygon orientation). This changes the sign of $W$ without changing $D$, $\rho$, or $\dot{\sigma}$.
The entropy production receives the same contribution from every edge:
\begin{equation}\label{eq:ring-sigma}\begin{aligned}
\dot{\sigma}&=(k_+-k_-)\ln\frac{k_+}{k_-}\\
&=\frac{2Db}{\rho^2h}\artanh b\\
&=\frac{2|W|}{\rho^2\sin\theta}
\artanh\!\left(\frac{|W|}{D}\tan\frac\theta2\right).
\end{aligned}\end{equation}
The last line is exactly the right-hand side of Eq.~\eqref{eq:observable-main}.
Thus the bound is attained, with no limiting procedure, for every strictly admissible set of input parameters.
It is precisely the regular polygon, the uniform edge currents, and the uniform edge biases that allow the geometric, Cauchy--Schwarz, and convexity steps of the general proof to saturate simultaneously.
The same homogeneous ring also saturates the sharp cycle-affinity bound of Ohga, Ito, and Kolchinsky \cite{OhgaItoKolchinsky2023}.

Figure~\ref{fig:scatter} tests the finite-support inequality on
415 randomly generated Markov networks, including models in which several
microscopic states share one observed value. On logarithmic axes, all
samples lie on or above the equality diagonal, in agreement with the
theoretical lower bound. The homogeneous-ring data saturate it, while the
other ensembles generally exhibit a positive gap. These observations
support the predicted validity and identify saturation as a special
geometric and dynamical feature, rather than a generic property of
random networks.

The coarse-valued examples are particularly informative because $d<N$,
while transitions between different observable levels still contribute
to the measured correlation slopes. Combining steady-state fluxes for
indistinguishable observed values preserves $D_a+D_b$, $\alpha_{ab}$,
and $\rho$. The log-sum inequality can reduce entropy production under
this aggregation, as expected from coarse-graining theory
\cite{Esposito2012,Bisker2017,Martinez2019}.
The stationary-flux reduction remains valid even when the observed
coarse-grained trajectory retains memory of hidden microscopic states.
It connects Eq.~\eqref{eq:observable-main} with the coarse-graining
framework of Esposito \cite{Esposito2012} and with path-based inference
methods \cite{VanderMeerErtelSeifert2022,HarunariDutta2022,VanderMeerDeguntherSeifert2023}.

\begin{figure}[t]
\centering
\includegraphics[width=\columnwidth]{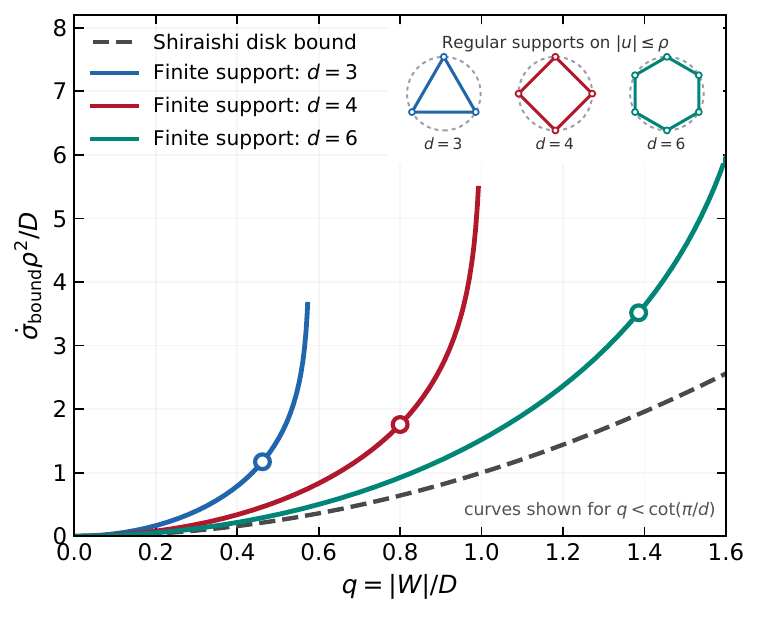}
\caption{Finite-support entropy-production bounds versus the normalized
correlation asymmetry $q=|W|/(D_a+D_b)$. The solid curves show
Eq.~\eqref{eq:observable-main} for $d=3,4,6$. The gray dashed curve is
Shiraishi's disk bound. The ordinate is
$\dot{\sigma}_{\mathrm{bound}}\rho^2/(D_a+D_b)$.
The open circles denote exact equality constructions for homogeneous
$d$-state rings with rates $k_+=1.8$ and $k_-=0.2$ and observable values
at the vertices of a regular $d$-gon on the unit circle.
The inset shows the three supports and their common enclosing-disk
geometry.
Each nonlinear curve is shown only
within its admissible domain $q<\cot(\pi/d)$.}
\label{fig:obs}
\end{figure}

\subsubsection{Connection with prior results}

\begin{figure*}[t]
\centering
\includegraphics[width=\textwidth]{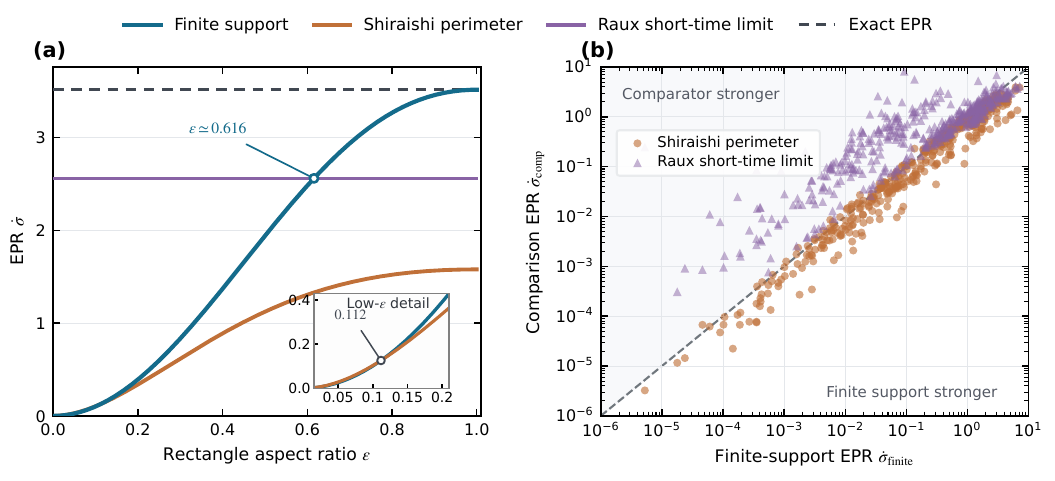}
\caption{\textbf{Complementarity of entropy-production bounds.}
(a) Exact entropy production rate (EPR) and three lower bounds for a
homogeneous four-state ring with $k_+=1.8$ and $k_-=0.2$ and observable
values $(a_i,b_i)=(1,\varepsilon),(-1,\varepsilon),
(-1,-\varepsilon),(1,-\varepsilon)$ in cyclic order.
Only the rectangle aspect ratio $\varepsilon$ varies, leaving the
underlying Markov dynamics and exact EPR unchanged.
The finite-support curve crosses Shiraishi's perimeter bound at
$\varepsilon\simeq0.112$ (inset) and the Raux short-time bound at
$\varepsilon\simeq0.616$.
(b) Pairwise comparison for the same 415 networks as in
Fig.~\ref{fig:scatter}. Each network contributes one orange circle at
$(\dot\sigma_{\mathrm{finite}},\dot\sigma_{\mathrm{per}})$ and
one purple triangle at
$(\dot\sigma_{\mathrm{finite}},\dot\sigma_{\mathrm{Raux}}^{0^+})$.
The dashed diagonal indicates equality. Points above or below it show
where the comparison bound or finite-support bound, respectively, is
stronger. In multicyclic networks the perimeter estimator uses
$L_{\max}$ from a fixed lexicographic greedy decomposition of positive
steady-state currents into uniform directed cycles.
The Raux estimator is calculated in the short-time limit
$\tau\to0^+$, with exact optimization over constant observable
shifts by weighted least squares.}
\label{fig:complementarity}
\end{figure*}

The finite-support result recovers Shiraishi's disk bound,
$\dot{\sigma}_{\mathrm{disk}}=4\alpha_{ab}^2/[\rho^2(D_a+D_b)]$
\cite{Shiraishi2023}, as $d\to\infty$ at fixed
$q=2|\alpha_{ab}|/(D_a+D_b)$.
For finite $d$, the improvement over the disk bound is
\begin{equation}\label{eq:gain}
\frac{\dot{\sigma}_{\mathrm{finite}}}{\dot{\sigma}_{\mathrm{disk}}}
=\sec^2(\pi/d)\frac{\artanh x}{x},\qquad
x=q\tan\frac{\pi}{d}.
\end{equation}
For small $x$, the enhancement approaches $\sec^2(\pi/d)$, while it
diverges as $x\to1^-$.  Its two ingredients are the regular-polygon
geometric factor of Ohga, Ito, and Kolchinsky
\cite{OhgaItoKolchinsky2023} and the full nonlinear relation between
stationary currents and their thermodynamic affinities retained here.
Figure~\ref{fig:obs} compares the finite-support lower bounds for three
support sizes with Shiraishi's disk bound. The nonlinear curves lie above
the disk estimate and become increasingly separated as the normalized
correlation asymmetry grows. The marked ring examples attain equality,
confirming that the stronger finite-support bound is sharp even away
from the weak-driving regime.
Shiraishi also derived a complementary perimeter bound
\cite{Shiraishi2023},
\begin{equation}\label{eq:shiraishi-perimeter}
\begin{aligned}
\dot{\sigma}_{\mathrm{per}}
&=\frac{16\pi^2\alpha_{ab}^2}
{(D_a+D_b)L_{\max}^2},\\
L_{\max}&=\max_{C\in\mathcal C_{\mathrm{ucd}}}
\sum_{(i,j)\in C}|u_j-u_i|,
\end{aligned}
\end{equation}
where $\mathcal C_{\mathrm{ucd}}$ is a uniform directed-cycle
decomposition consistent with the stationary currents.
Combining this estimate with Eq.~\eqref{eq:gain} gives
\begin{equation}\label{eq:perimeter-gain}
\frac{\dot{\sigma}_{\mathrm{finite}}}
{\dot{\sigma}_{\mathrm{per}}}
=\left(\frac{L_{\max}}{2\pi\rho}\right)^2
\sec^2\!\left(\frac{\pi}{d}\right)
\frac{\artanh x}{x}.
\end{equation}
Thus the relative strength depends on the dimensionless cycle
perimeter $L_{\max}/\rho$: the finite-support estimate is stronger
precisely when
$L_{\max}>2\pi\rho\cos(\pi/d)\sqrt{x/\artanh x}$.
For a homogeneous ring whose observable values form a regular
$d$-gon, $L_{\max}=2d\rho\sin(\pi/d)$, yielding
\begin{equation}\label{eq:regular-perimeter-gain}
\frac{\dot{\sigma}_{\mathrm{finite}}}
{\dot{\sigma}_{\mathrm{per}}}
=\left(\frac{d\tan(\pi/d)}{\pi}\right)^2
\frac{\artanh x}{x}>1.
\end{equation}
The first factor is the finite-polygon correction to the
isoperimetric geometry. The second retains the nonlinear
current--affinity relation.  There is nevertheless no universal
ordering for general observable geometries.  For example, in the
weak-driving limit of a four-state ring whose observed values form
an increasingly thin rectangle,
$\dot{\sigma}_{\mathrm{per}}/\dot{\sigma}_{\mathrm{disk}}
\to\pi^2/4$, while
$\dot{\sigma}_{\mathrm{finite}}/\dot{\sigma}_{\mathrm{disk}}
\to2$.
The perimeter bound uses additional information about the
current-carrying cycles and can therefore be stronger than our
finite-support bound in such geometries.

Raux, Maier, and Seifert have developed a complementary
correlation-based approach to entropy-production inference
\cite{RauxMaierSeifert2026}.
Their bound applies at arbitrary correlation lag in stationary
processes, while a short-time formulation is also available
for time-dependent dynamics.
In our notation, the stationary short-time bound reads
$\dot{\sigma}\geq 2W^2/K_{ab}$, where $W=2\alpha_{ab}$ and
\begin{equation}
K_{ab}=
\left.\partial_\tau\left[
C_{a^2,b^2}(\tau)+C_{b^2,a^2}(\tau)
-2C_{ab,ab}(\tau)
\right]\right|_{\tau=0^+}.
\end{equation}
Thus, their estimator exploits short-time
correlations involving products of the observables.
Optimizing over constant shifts of the observables further
tightens the bound and yields an estimate no weaker than
Shiraishi's disk bound.
By contrast, our finite-support bound retains the
inputs $D_a$, $D_b$, $\alpha_{ab}$, and $\rho$ entering
Shiraishi's disk estimate, supplemented by the
static information that the observable pair takes $d$
distinct values.
The two approaches therefore exploit different types of
observable information. 

Figure~\ref{fig:complementarity} tests whether one EPR bound
uniformly dominates the others. Panel (a) varies the observable
geometry while holding the dynamics and exact EPR fixed. The two
crossings provide analytic examples in which the finite-support bound
is stronger or weaker than either reference bound.
In particular, sufficiently thin rectangles favor the perimeter and
Raux short-time bounds, whereas the square saturates our bound and
places it above both alternatives.
Panel (b) demonstrates both pairwise orderings in broader network
ensembles. The finite-support estimate is stronger than the perimeter
estimate in 337 of 415 samples and weaker in 78. It is stronger than
the shift-optimized Raux short-time estimate in 99 samples and weaker
in 316. These frequencies depend on the chosen ensemble and, for the
perimeter bound, on the cycle decomposition. They are not universal
probabilities. The comparison illustrates the distinct information
used by the methods.
There is
no universal pairwise ordering between our bound and either of these
two reference estimates.

The dynamical-activity bounds of Gu \cite{Gu2024} retain complementary
information about the total transition traffic.
The same distinction between entropy production and dynamical activity
underlies kinetic and entropic perturbation inequalities
\cite{DechantSasa2021,DiTerlizzi2024,DiTerlizzi2025}.
The recent pathwise calculus for Markov jumps
\cite{DieballStutzerGodec2026} provides a broader setting for
trajectory-dependent fluctuation relations, while the sharp bound
in Eq.~\eqref{eq:observable-main} uses only the initial correlation slopes
of state observables.
Section~\ref{subsec:finite-time} gives a finite-lag counterpart.

\subsection{Dissipation and coherence of oscillatory modes}\label{subsec:spectral}

We now present a corollary of the main bound.
An oscillatory relaxation mode can be represented by a complex state-observable profile $v$ satisfying
\begin{equation}\label{eq:mode-eigenvalue}
R^{\mathsf T}v=(-r+\mathrm{i}\omega)v,\qquad r>0,\ \omega>0.
\end{equation}
Because $R$ acts on state probabilities and $R^{\mathsf T}$ acts on state observables, this is a left eigenvector of the transition rate matrix $R$. Both matrices have the same eigenvalues. The decay rate $r$ sets the lifetime of the mode, while $\omega$ sets its oscillation frequency.
We write the coherence ratio as $\mathcal R=\omega/r$ (to distinguish it from the transition rate matrix $R$), the stationary mean-square mode amplitude $V=\sum_ip_i^{\rm ss}|v_i|^2$, and the entropy produced in one oscillation period $\Delta S=2\pi\dot{\sigma}/\omega$.

The connection with the observable problem is immediate once $\operatorname{Re}v$ and $\operatorname{Im}v$ are treated as the two measured variables.
Stationarity gives the quadratic-form identity
\begin{equation}\label{eq:mode-quadratic-main}
\sum_ip_i^{\rm ss}\overline v_i(R^{\mathsf T}v)_i
=-D(v)+\mathrm{i}W(v).
\end{equation}
The eigenvalue equation gives the same expression as $(-r+\mathrm{i}\omega)V$.
Matching its real and imaginary parts shows that
\begin{equation}\label{eq:mode-DW-main}
D(v)=rV,\qquad W(v)=\omega V,
\qquad \frac{|W(v)|}{D(v)}=\mathcal R.
\end{equation}
Thus the asymmetry of the short-time correlations of an eigenfunction is exactly its frequency-to-decay ratio.

Let $\rho(v)$ be the smallest enclosing-disk radius of the complex numbers $v_i$.
The disk factor
\begin{equation}\label{eq:mode-eta}
\etad=\frac{V}{\rho(v)^2},\qquad 0<\etad\leq1,
\end{equation}
compares the average squared amplitude with the squared radius required to contain all mode values.
The upper bound follows from the zero stationary mean of a nonstationary eigenfunction.
Substituting Eq.~\eqref{eq:mode-DW-main} into the observable result, and multiplying by the period $2\pi/\omega$, yields
\begin{equation}\label{eq:spectral-main}
\Delta S\geq\etad \frac{4\pi}{\sin(2\pi/d)}\artanh\!\left(\mathcal R\tan\frac{\pi}{d}\right).
\end{equation}
The derivation of Eq.~\eqref{eq:spectral-main}, including the bound on $\etad$ and its equality condition, is given in Appendix~\ref{app:mode}.
Here $d$ is the number of distinct complex values $v_i$, including a zero value if one occurs.
For an oscillatory eigenmode, $0<\mathcal R<\cot(\pi/d)$ is automatic from the observable geometry.
The factor $\etad$ is essential: the pair $(r,\omega)$ by itself does not specify how eigenfunction amplitudes are distributed over microscopic states.
In particular, $\etad=1$ if and only if all $|v_i|$ are equal.

Equation~\eqref{eq:spectral-main} unifies the observable and spectral descriptions through the substitution $q=\mathcal R$.
Compared with the linear dissipation bound $2\pi\etad\mathcal R$, the right-hand side is enhanced by the factor in Eq.~\eqref{eq:gain}, evaluated at $x=\mathcal R\tan(\pi/d)$.
Gu's earlier bound reads $\Delta S\geq2\pi\eta\mathcal R$ with $\eta=V/\max_i|v_i|^2$ \cite{Gu2026}.
Since $\eta\leq\etad$, the disk factor and finite polygon both improve that result.
For fixed $\mathcal R$ and increasing $d$, Eq.~\eqref{eq:spectral-main} approaches the linear disk bound.
When an eigenmode has constant modulus, $\etad=1$ and the full nonlinear homogeneous-ring curve applies to arbitrary network topology, even without translational symmetry.
The uniformly driven ring reaches equality in the full nonlinear curve, reproducing the homogeneous-ring calculation of Oberreiter, Seifert and Barato \cite{OberreiterSeifertBarato2022}.

The limiting and special cases separate three physical ingredients: the number of distinct mode values, the radial variation of an oscillatory mode, and the degree of nonequilibrium circulation.
The first is also constrained by the cycle-affinity geometry of Barato and Seifert and Ohga and collaborators \cite{BaratoSeifert2017,OhgaItoKolchinsky2023}.
The radial factor becomes relevant when a mode is localized on a subset of states, as emphasized in Gu's spectral analysis and Shiraishi's counterexample \cite{Gu2026,Shiraishi2026}.
For weak-noise stochastic limit cycles, phase diffusion and the thermodynamic uncertainty relation lead to coherence bounds under different assumptions \cite{SantolinFalasco2025,NagayamaIto2025,Kolchinsky2026}.
These weak-noise results and the present finite-state spectral result concern distinct model classes and definitions of coherence, so their domains of validity are consistent.

Even for uniform stationary probabilities, the disk factor need not equal
unity: $V$ averages $|v_i|^2$ over all states, whereas $\rho(v)$ also
reflects the extreme observable values.  A mode concentrated on a subset
of states can therefore have $\etad<1$ even when each state has the same
stationary weight.  Equation~\eqref{eq:spectral-main} retains this physical
information instead of assuming that the decay rate and oscillation frequency
alone determine the dissipative cost.

\subsection{Finite-time extension}\label{subsec:finite-time}

The initial slopes in Eq.~\eqref{eq:observable-main} may be difficult to
estimate when the sampling interval is not very short.  For a stationary
Markov jump process, a related dissipation bound can be obtained directly
from two-time correlations at any lag $\tau>0$.  With the notation of
Eq.~\eqref{eq:correlation-slopes}, define the correlation asymmetry and
autocorrelation losses by
\begin{equation}\label{eq:finite-time-observables}\begin{aligned}
\chi_{ab}(\tau)&=C_{ab}(\tau)-C_{ba}(\tau),\\
\Delta_a(\tau)&=C_{aa}(0)-C_{aa}(\tau),\\
\Delta_b(\tau)&=C_{bb}(0)-C_{bb}(\tau).
\end{aligned}\end{equation}
For nonconstant observable pairs, $\Delta_a(\tau)+\Delta_b(\tau)>0$.
The number $d$ of distinct joint observable values and their enclosing-disk
radius $\rho$ are unchanged.  Set
$q_\tau=|\chi_{ab}(\tau)|/[\Delta_a(\tau)+\Delta_b(\tau)]$.
When $\chi_{ab}(\tau)\neq0$, necessarily $d\geq3$ and
$0<q_\tau<\cot(\pi/d)$.  The entropy production satisfies
\begin{equation}\label{eq:finite-time-main}
\dot{\sigma}\geq
\frac{2|\chi_{ab}(\tau)|}{\tau\rho^2\sin(2\pi/d)}
\artanh\!\left(
q_\tau\tan\frac{\pi}{d}
\right).
\end{equation}
For $\chi_{ab}(\tau)=0$, the right-hand side is understood to be zero.
Equation~\eqref{eq:finite-time-main} is valid for every positive
observation lag, not only in a short-time approximation.
Its derivation using an auxiliary process and relative-entropy contraction to endpoint statistics is given in Appendix~\ref{app:finite-time}.

To understand its origin, let $P(\tau)=e^{R\tau}$ be the transition
probability matrix.  The auxiliary transition rate matrix
$R^{(\tau)}=[P(\tau)-I]/\tau$ has the same stationary distribution and
observable values as the original process.  Its initial autocorrelation
decay rates are $\Delta_a(\tau)/\tau$ and $\Delta_b(\tau)/\tau$,
while its fluctuation oscillation is $\chi_{ab}(\tau)/(2\tau)$.
Equation~\eqref{eq:observable-main} therefore applies to this auxiliary
process.  
Its entropy production cannot exceed that of the original
process, by the data-processing inequality for path-space
relative entropy: retaining only the initial and final
states discards information about the time-reversal
asymmetry of the intervening trajectory.
This endpoint coarse-graining principle is also central
to the finite-time correlation bounds of Ref.~\cite{VanVuVoSaito2024}.

The short-lag limits
$\chi_{ab}(\tau)=2\alpha_{ab}\tau+O(\tau^2)$ and
$\Delta_a(\tau)+\Delta_b(\tau)=(D_a+D_b)\tau+O(\tau^2)$
show that Eq.~\eqref{eq:finite-time-main} reduces exactly to
Eq.~\eqref{eq:observable-main} as $\tau\to0^+$.
At fixed $\tau$, a quadratic disk bound follows from
Eq.~(13) of Van Vu, Vo, and Saito
\cite{VanVuVoSaito2024}, after optimizing over constant
shifts of the observables:
\begin{equation}
\dot{\sigma}_{\mathrm{disk}}^{(\tau)}
=
\frac{\chi_{ab}(\tau)^2}
{\tau\rho^2[\Delta_a(\tau)+\Delta_b(\tau)]}.
\end{equation}
Relative to this disk-radius bound, our nonlinear
finite-support result provides an enhancement factor
of precisely the form in Eq.~\eqref{eq:gain}, with
$x$ replaced by $x_\tau=q_\tau\tan(\pi/d)$.
We emphasize that Ref.~\cite{VanVuVoSaito2024}
also includes a complementary cycle-perimeter bound.
This comparison therefore does not imply a uniform
improvement over its complete finite-time inequality.

Unlike the short-time result, Eq.~\eqref{eq:finite-time-main}
is not claimed to be sharp at a prescribed nonzero $\tau$.
The observation interval may contain multiple jumps,
including complete cycles that cannot be resolved from
its endpoints.
Related finite-time constraints have been obtained by
Liang and Pigolotti, who bound reversal asymmetry through
the largest cycle affinity
\cite{LiangPigolotti2023}.
Raux, Maier, and Seifert have also derived finite-lag
entropy-production bounds based on correlation asymmetry
and additional correlations involving products of the
observables \cite{RauxMaierSeifert2026}.
Our finite-time construction   provides a complementary constraint
based on different observable information.

\section{Discussion and Outlook}\label{sec:discussion}

Our central result identifies the minimum stationary entropy production compatible with a restricted set of observable information: the combined initial autocorrelation decay, the initial cross-correlation asymmetry, and the number $d$ and minimum enclosing-disk radius $\rho$ of distinct joint observable values. The resulting nonlinear lower bound is sharp because a uniformly driven ring attains equality for every admissible nonzero asymmetry. It applies even when different microscopic states produce the same readout. Finite observable support therefore supplies thermodynamic information that is absent from the enclosing-disk radius alone.

The spectral corollary offers a complementary interpretation of the same geometry. The factor $\etad=V/\rho(v)^2$ compares the stationary mean-square amplitude of an oscillatory eigenmode with the squared radius of its smallest enclosing disk. Consequently, a mode's frequency-to-decay ratio does not by itself capture the dependence of the dissipation bound on its spatial profile. Uniform rings saturate the nonlinear spectral inequality, but whether it is optimal for modes with $\etad<1$ remains an open question.

For experimental inference, the allowed \emph{joint} observable values must be distinguished from the levels visited in a finite record. Unvisited values may lead to underestimates of $d$ or $\rho$, while noisy state identification complicates both support calibration and the estimation of correlation slopes \cite{VanderMeerSaito2025,GhosalBisker2026}. Fluorescence correlation spectroscopy and single-molecule energy-transfer measurements motivate correlation-based analysis, although applying a finite-support bound requires a justified discrete readout model \cite{Elson2011,Nettels2024}. Promising directions include uncertainty-aware inference from noisy and finitely sampled trajectories, determining sharp bounds at nonzero lag, and combining support geometry with higher-order correlation statistics without reconstructing hidden transitions.

\begin{acknowledgments}
GPT-6 Pro and Claude Opus 5.5 were used for literature search, computations and checking proofs.
The authors take full responsibility for all the contents.
This work is part of \href{https://github.com/the-omega-institute/trureturing}{Trureturing}, an open research project on AI-assisted mathematics.

\end{acknowledgments}

\appendix
\section{Correlation slopes and stationary currents}\label{app:slopes}

We show how the short-time correlation measurements are related to transition rates and steady-state fluxes.
For any real state functions $f$ and $g$, the master equation gives
\begin{equation}\label{eq:correlation-derivative}
C'_{fg}(0^+)=\sum_ip_i^{\rm ss} f_i(R^{\mathsf T}g)_i.
\end{equation}
In particular, use the identity
\begin{equation}\label{eq:squares-identity}
2f_i(f_j-f_i)=(f_j^2-f_i^2)-(f_j-f_i)^2.
\end{equation}
After multiplication by $p_i^{\rm ss}R_{ji}$ and summation over directed transitions, the first term on the right vanishes because $\sum_ip_i^{\rm ss}(R^{\mathsf T}f^2)_i=0$.
Thus
\begin{equation}\label{eq:self-slopes}
C'_{ff}(0^+)
=-\frac12\sum_{i\to j}A_{ij}(f_j-f_i)^2.
\end{equation}
Taking $f=a$ and $f=b$ gives the autocorrelation decay rates $D_a$ and $D_b$ in Eq.~\eqref{eq:DW-def}.

For the antisymmetric part, expand Eq.~\eqref{eq:correlation-derivative} for $f=a,g=b$ and for $f=b,g=a$.
The terms proportional to $a_ib_i$ cancel, leaving
\begin{equation}\label{eq:cross-slope-proof}
\begin{aligned}
C'_{ab}(0^+)-C'_{ba}(0^+)
&=\sum_{i\to j}(F_{ij}-F_{ji})\\
&\quad\times(a_i b_j-b_i a_j)\\
&=\sum_{i\to j}J_{ji}^{\rm ss}
\operatorname{Im}(\overline u_i u_j)=W.
\end{aligned}
\end{equation}
Finally, stationarity implies $\sum_j(F_{ij}-F_{ji})=0$ at each state.
Multiplying by $u_i$ and summing over $i$ gives the useful current-conservation identity
\begin{equation}\label{eq:translation-invariance}
\sum_{i\to j}J_{ji}^{\rm ss}(u_j-u_i)=0.
\end{equation}
Indeed, under $u_i\mapsto u_i-c$ the quantity
$\operatorname{Im}(\overline u_i u_j)$ changes by
$-\operatorname{Im}[\overline c(u_j-u_i)]$.
Equation~\eqref{eq:translation-invariance} therefore proves the claimed translation invariance of $W$. Translation invariance of $D$ is immediate from its definition.

\section{Geometry and derivation of the observable bound}\label{app:observable}

The physical ingredients are steady-state current conservation, the geometry of the observable values, and the entropy production needed to maintain the currents. Here $D=D_a+D_b$ and $W=2\alpha_{ab}$ are only abbreviations.
We first merge states with identical observable values, then decompose the remaining stationary currents into directed cycles.
Two complementary area estimates constrain each cycle, and convexity combines the resulting cyclewise entropy costs into a bound on the full system.
Importantly, the argument uses only stationary fluxes: no Markov assumption is imposed on the observed coarse-grained trajectory.

\subsection{Reduction to distinct observable values}

Partition the microscopic state space into classes $A,B,\ldots$, with $i$ and $j$ in the same class precisely when $u_i=u_j$.
For each class, define
\begin{equation}\label{eq:coarse-fluxes}\begin{aligned}
\overline p_A^{\rm ss}&=\sum_{i\in A}p_i^{\rm ss},\\
\overline F_{AB}&=\sum_{i\in A,j\in B}F_{ij},\\
\overline R_{BA}&=\overline F_{AB}/\overline p_A^{\rm ss}
\qquad(A\ne B).
\end{aligned}\end{equation}
The outgoing-minus-incoming flux vanishes for each class because it vanishes after summing the microscopic stationarity equations for its members.
Consequently, the $\overline R_{BA}$ define a stationary $d$-state chain with distribution $\overline p^{\rm ss}$.
Irreducibility and bidirectionality are inherited from the original connected network.

An edge joining states within the same class has $u_i=u_j$ and contributes neither to $D$ nor to $W$.
For any two different classes, the displacement $u_B-u_A$ and the oriented-area factor $\operatorname{Im}(\overline u_Au_B)$ do not depend on which microscopic states are chosen.
Summing their fluxes therefore leaves $D$ and $W$ unchanged. The set of distinct values, and hence $\rho$, is also unchanged.

The entropy production can decrease, which is the direction needed for a lower bound.
Indeed, for the microscopic flux pairs connecting $A$ and $B$, the log-sum inequality applied once to each orientation gives
\begin{equation}\label{eq:logsum-coarse}\begin{aligned}
&\sum_{i\in A,j\in B}
(F_{ij}-F_{ji})\ln\frac{F_{ij}}{F_{ji}}\\
&\quad\geq(\overline F_{AB}-\overline F_{BA})
\ln\frac{\overline F_{AB}}{\overline F_{BA}}.
\end{aligned}\end{equation}
The sums involving flux ratios are understood over allowed microscopic edges. Absent edges carry no flux and no entropy contribution.
All intraclass entropy contributions are nonnegative and can be dropped.
It follows that
\begin{equation}\label{eq:coarse-ineq}
\dot{\sigma}\geq\overline{\dot{\sigma}},
\qquad \overline D=D,\quad \overline W=W.
\end{equation}
We may therefore prove the desired inequality for this effective $d$-state chain and carry it back to the microscopic system using Eq.~\eqref{eq:coarse-ineq}.
In the remaining calculation, all quantities refer to this chain unless otherwise indicated.

Let $c_*$ be a center of its smallest enclosing disk.
For $W\ne0$, the radius is positive, so we can introduce normalized values
\begin{equation}\label{eq:z-normalization}
z_i=\frac{u_i-c_*}{\rho},\qquad |z_i|\leq1.
\end{equation}
The translation and scaling properties in Appendix~\ref{app:slopes} imply
\begin{equation}\label{eq:normalized-DW}
D(z)=D/\rho^2,\qquad W(z)=W/\rho^2.
\end{equation}
This normalization allows both geometric estimates below to be stated for the unit disk, without changing the dimensionless ratio $W/D$.

\subsection{Cycle decomposition and allocation of entropy}

Reorient each edge with a nonzero steady-state current so that its current $J_e>0$, where $J_e$ denotes $J_{ji}^{\rm ss}$ along that oriented edge, and define the dimensionless current bias
\begin{equation}\label{eq:edge-bias}
b_e=\frac{J_e}{A_e}\in(0,1).
\end{equation}
Both fluxes are strictly positive, hence $J_e<A_e$.
The contribution of this edge to entropy production can be written exactly as
\begin{equation}\label{eq:exact-edge-entropy}\begin{aligned}
\dot{\sigma}_e&=J_e\ln\frac{F_{ij}}{F_{ji}}
=J_e\ln\frac{A_e+J_e}{A_e-J_e}\\
&=2J_e\artanh b_e.
\end{aligned}\end{equation}
No expansion in the bias $b_e$ is made. This is where the eventual nonlinear improvement enters.

In a steady state, probability cannot accumulate at any state.
Every positive edge therefore belongs to a directed cycle: following outgoing positive-current edges cannot terminate, and in a finite graph a vertex must repeat.
Subtracting the smallest current on such a cycle and repeating yields a finite decomposition
\begin{equation}\label{eq:current-decomposition}
J_e=\sum_{C:\,e\in C}j_C,\qquad j_C>0,
\end{equation}
into simple directed cycles $C$.
There are no two-vertex directed cycles because each undirected edge has only one positive-current orientation.
Thus every cycle contains $n\geq3$ distinct vertices, and since there are $d$ aggregated states, $n\leq d$.

For a cycle $C$, let $\ell_e=|z_j-z_i|$ and define
\begin{equation}\label{eq:polygon}\begin{aligned}
P_C&=\sum_{e\in C}\ell_e,\\
A_C&=\sum_{(i\to j)\in C}
\operatorname{Im}(\overline z_i z_j).
\end{aligned}\end{equation}
The quantity $P_C$ is its polygonal perimeter, and $A_C$ is twice its signed area, whether the polygon is simple in the geometric sense or self-intersecting.
Notice that directed cycles in the state graph need not trace non-self-intersecting polygons in the complex plane.

Allocate to $C$ the following parts of $D(z)$, $W(z)$, and $\overline{\dot{\sigma}}$:
\begin{equation}\label{eq:cycle-allocations}\begin{aligned}
x_C&=\frac{j_C}{2}\sum_{e\in C}\frac{\ell_e^2}{b_e},\\
y_C&=j_CA_C,\\
s_C&=2j_C\sum_{e\in C}\artanh b_e.
\end{aligned}\end{equation}
To see the first formula, use $A_e=J_e/b_e$ in the definition of $D(z)$ and insert Eq.~\eqref{eq:current-decomposition}. Each cycle then receives the share proportional to $j_C$.
The same current decomposition turns the oriented-area sum for $W(z)$ into $\sum_C j_CA_C$.
Finally, Eq.~\eqref{eq:exact-edge-entropy} shows that the shares $s_C$ add to the positive-current part of the total entropy.

Edges with $J_e=0$ do not appear in the cycle decomposition.
Each such edge contributes a separate term $x_e=A_e\ell_e^2/2$ and $y_e=s_e=0$.
Including those terms in all subsequent sums gives the exact bookkeeping identities
\begin{equation}\label{eq:cycle-sum}
\sum x=\frac{D}{\rho^2},\qquad
\sum y=\frac{W}{\rho^2},\qquad
\sum s=\overline{\dot{\sigma}}.
\end{equation}
Thus a bound on $s_C$ in terms of $x_C$ and $y_C$ will translate into a bound involving only the measured quantities $D$ and $W$.

\subsection{Two geometric bounds on each cycle}

Let a cycle have $n$ vertices and perimeter $P_C$.
The polygon isoperimetric inequality gives
\begin{equation}\label{eq:polygoniso}
|A_C|\leq\frac{P_C^2}{2n\tan(\pi/n)}.
\end{equation}
This is the usual regular-polygon optimal constant for twice the area. It also applies to self-intersecting polygons \cite[Supplemental Material, Sec.~B.5, Eq.~(S19)]{OhgaItoKolchinsky2023}.
It encodes the constraint imposed by the finite number of distinct observable values, but it does not yet use their containment in the unit disk.

A second inequality incorporates this containment.
For each edge write $m=(z_i+z_j)/2$ and $\delta=z_j-z_i$, with $\ell_e=|\delta|$.
By expanding the squares of its endpoints, one obtains
\begin{equation}\label{eq:edgegeo}\begin{aligned}
|m|^2+\frac{\ell_e^2}{4}
&=\frac{|z_i|^2+|z_j|^2}{2}\leq1,\\
\left|\operatorname{Im}(\overline z_i z_j)\right|
&=\left|\operatorname{Im}(\overline m\,\delta)\right|\\
&\leq \ell_e\sqrt{1-\ell_e^2/4}.
\end{aligned}\end{equation}
The first line bounds the possible distance of an edge midpoint from the disk center. The second bounds that edge's contribution to oriented area.
Summing over the cycle and applying Cauchy--Schwarz gives
\begin{equation}\label{eq:area-seq}\begin{aligned}
|A_C|^2
&\leq\left(\sum_e\ell_e
\sqrt{1-\ell_e^2/4}\right)^2\\
&\leq P_C\left(P_C-\frac14\sum_e\ell_e^3\right)\\
&\leq P_C^2
\left(1-\frac{P_C^2}{4n^2}\right).
\end{aligned}\end{equation}
The final step uses $\sum_e\ell_e^3\geq P_C^3/n^2$, which follows from convexity of $x\mapsto x^3$ on $x\geq0$.
Taking the square root yields
\begin{equation}\label{eq:disk-area}
|A_C|\leq P_C\sqrt{1-P_C^2/(4n^2)}.
\end{equation}
The two estimates are complementary: Eq.~\eqref{eq:polygoniso} is effective for relatively short perimeters, while Eq.~\eqref{eq:disk-area} becomes the stronger constraint when the polygonal perimeter is large relative to the disk radius.

\subsection{From cycle geometry to nonlinear entropy}

Consider a cycle with $A_C\ne0$ and abbreviate
\begin{equation}\label{eq:cycle-parameters}
X=\frac{x_C}{j_C}
=\frac12\sum_{e\in C}\frac{\ell_e^2}{b_e},\quad
q_C=\frac{|A_C|}{X},\quad
\alpha=\frac{\pi}{n}.
\end{equation}
Here $q_C=|y_C|/x_C$ is the absolute asymmetry allocated to this cycle.
Weighted Cauchy--Schwarz gives
\begin{equation}\label{eq:weighted-cs}
P_C^2=\left(\sum_e\ell_e\right)^2
\leq\left(\sum_e\frac{\ell_e^2}{b_e}\right)
\left(\sum_e b_e\right)
=2X\sum_e b_e.
\end{equation}
Combining this with Eq.~\eqref{eq:polygoniso} implies
\begin{equation}\label{eq:cycle-domain}
q_C\leq
\left(\frac1n\sum_e b_e\right)\cot\alpha
<\cot\alpha.
\end{equation}
The strict inequality follows from $0<b_e<1$.
It ensures in particular that $q_C\tan\alpha<1$, the domain needed for the inverse hyperbolic tangent below.

Next, convexity and monotonicity of $\artanh$ on $[0,1)$ give
\begin{equation}\label{eq:cycle-entropy-jensen}\begin{aligned}
\frac{s_C}{j_C}
&=2\sum_e\artanh b_e\\
&\geq2n\artanh\!\left(\frac1n\sum_e b_e\right)\\
&\geq2n\artanh\!\left(\frac{P_C^2}{2nX}\right).
\end{aligned}\end{equation}
The first inequality is Jensen's inequality. The second uses Eq.~\eqref{eq:weighted-cs}.
Thus the cycle entropy cannot be reduced arbitrarily by redistributing its current biases among different edges.

Introduce dimensionless perimeter and area variables
\begin{equation}\label{eq:ta-def}
t=\frac{P_C}{2n},\qquad
 a=\frac{|A_C|}{2n}.
\end{equation}
Every edge is a chord of the unit disk, so $0<t<1$ for a nondegenerate cycle with at least three distinct vertices.
The two area bounds become
\begin{equation}\label{eq:case-key}
a\leq\min\{t^2\cot\alpha,
 t\sqrt{1-t^2}\}.
\end{equation}
Moreover,
$P_C^2/(2nX)=q_Ct^2/a$ and $|y_C|=2nj_Ca$.
Consequently Eq.~\eqref{eq:cycle-entropy-jensen} reads
\begin{equation}\label{eq:entropy-vs-area}
\frac{s_C}{|y_C|}\geq
\frac1a\artanh\!\left(\frac{q_Ct^2}{a}\right).
\end{equation}
At fixed positive $q_C$ and $t$, the expression on the right decreases as $a$ increases (both $1/a$ and its $\artanh$ argument decrease).
We can therefore replace $a$ by either geometric upper bound to obtain a valid lower estimate.
The two area bounds cross at $t=\sin\alpha$, which explains the following division into cases.

{Case 1: $t\leq\sin\alpha$.}
Using $a\leq t^2\cot\alpha$ gives
\begin{equation}\label{eq:case-short}\begin{aligned}
\frac{s_C}{|y_C|}
&\geq\frac{\tan\alpha}{t^2}
\artanh(q_C\tan\alpha)\\
&\geq\frac{2}{\sin(2\alpha)}
\artanh(q_C\tan\alpha).
\end{aligned}\end{equation}
The final inequality is $t^2\leq\sin^2\alpha$ together with
$\tan\alpha/\sin^2\alpha=2/\sin(2\alpha)$.

{Case 2: $t\geq\sin\alpha$.}
Set $\beta=\arcsin t\geq\alpha$.
The disk-area constraint $a\leq t\sqrt{1-t^2}$ yields
\begin{equation}\label{eq:beta}
\frac{s_C}{|y_C|}\geq
\frac{2}{\sin(2\beta)}
\artanh(q_C\tan\beta).
\end{equation}
This lower bound has exactly the same functional form as in Case 1, but with $\beta$ in place of $\alpha$.
To compare them, define for $0<\phi<\pi$ the even function
\begin{equation}\label{eq:Hdef}\begin{aligned}
H_\phi(q)&=\frac{2q}{\sin\phi}
\artanh\!\left(q\tan\frac\phi2\right)\\
&=\sec^2\!\frac\phi2
\sum_{k=0}^\infty
\frac{q^{2k+2}\tan^{2k}(\phi/2)}{2k+1},
\end{aligned}\end{equation}
where $|q|<\cot(\phi/2)$.
Every term in the convergent power series is nonnegative, and its coefficient is nondecreasing in $\phi$.
Thus $H_\phi(q)$ is nondecreasing in $\phi$ wherever both sides are defined.
In Case 2, Eq.~\eqref{eq:entropy-vs-area} ensures $q_C\tan\beta<1$, so the comparison with $\beta\geq\alpha$ is legitimate.
Multiplying Eqs.~\eqref{eq:case-short} and \eqref{eq:beta} by $|y_C|=x_Cq_C$, and using this monotonicity in the second case, we obtain in both cases
\begin{equation}\label{eq:percycle}\begin{aligned}
s_C&\geq x_C H_{2\pi/n}(y_C/x_C)\\
&\geq x_C H_{2\pi/d}(y_C/x_C).
\end{aligned}\end{equation}
The sign of $A_C$ does not matter because $H_\phi$ is even.
The second inequality follows from $n\leq d$ and the same monotonicity in $\phi$.
For cycles with $A_C=0$ or edges with $J_e=0$, the right-hand side vanishes since $H_\phi(0)=0$, and the inequality remains true.

\subsection{Combining cycles: the global inequality}

For the fixed angle $\phi=2\pi/d$, the function $H_\phi$ is strictly convex on its domain.
Indeed, differentiating the expression above twice gives
\begin{equation}\label{eq:convexity}
H_\phi''(q)=
\frac{4\tan(\phi/2)}
{\sin\phi[1-q^2\tan^2(\phi/2)]^2}>0.
\end{equation}
For every term in Eq.~\eqref{eq:cycle-sum} with $x>0$, let $w=x/(D/\rho^2)$ and $q=y/x$.
The weights sum to one, and their weighted mean asymmetry is exactly
\begin{equation}\label{eq:jensen-mean}
\sum wq=\frac{\sum y}{\sum x}=\frac WD.
\end{equation}
By Eq.~\eqref{eq:cycle-domain}, each $q$ is in
$(-\cot(\pi/d),\cot(\pi/d))$. Zero-current terms have $q=0$.
Hence Jensen's inequality can be applied without leaving that domain:
\begin{equation}\label{eq:jensen}\begin{aligned}
\dot{\sigma}&\geq\overline{\dot{\sigma}}=\sum s\\
&\geq\sum x H_{2\pi/d}(y/x)\\
&\geq\frac{D}{\rho^2}
H_{2\pi/d}(W/D).
\end{aligned}\end{equation}
Substituting the definition of $H$ reproduces Eq.~\eqref{eq:observable-main}, with an absolute value of $W$ because $H$ is even.
The same weighted-average argument, combined with $|q|<\cot(\pi/d)$ for every cycle, proves the strict global range $|W|/D<\cot(\pi/d)$ when $W\ne0$.
This completes the proof.

\section{Oscillatory-mode reduction and disk geometry}\label{app:mode}

Let $v$ satisfy Eq.~\eqref{eq:mode-eigenvalue} with eigenvalue
$\lambda=-r+\mathrm{i}\omega\ne0$.
Stationarity immediately gives
\begin{equation}\label{eq:zero-mean}
0=\sum_ip_i^{\rm ss}(R^{\mathsf T}v)_i
=\lambda\sum_ip_i^{\rm ss} v_i,
\qquad\sum_ip_i^{\rm ss} v_i=0.
\end{equation}
This zero-mean property is special to nonstationary eigenfunctions and will be needed to bound the disk factor.

To prove the quadratic-form identity explicitly, start from
\begin{equation}\label{eq:quadform-expansion}
\sum_ip_i^{\rm ss}\overline v_i(R^{\mathsf T}v)_i
=\sum_{i\to j}\!\bigl[
F_{ij}\overline v_i(v_j-v_i)
+F_{ji}\overline v_j(v_i-v_j)\bigr].
\end{equation}
The imaginary part of the bracket is
$J_{ji}^{\rm ss}\operatorname{Im}(\overline v_i v_j)$.
Its real part is
\begin{equation}\label{eq:real-quadratic-edge}
-\frac{A_{ij}}2|v_j-v_i|^2
+\frac{J_{ji}^{\rm ss}}2(|v_j|^2-|v_i|^2).
\end{equation}
The last term vanishes after summing over all edges, by stationary current conservation [Eq.~\eqref{eq:translation-invariance}, applied to the real state function $|v|^2$].
Combining real and imaginary parts recovers
\begin{equation}\label{eq:quadform}
\sum_ip_i^{\rm ss}\overline v_i(R^{\mathsf T}v)_i
=-D(v)+\mathrm{i}W(v).
\end{equation}
Using $R^{\mathsf T}v=\lambda v$, its left-hand side is
$\lambda V=(-r+\mathrm{i}\omega)V$.
Since $V>0$, comparison of the two parts establishes
\begin{equation}\label{eq:DW-mode}
D(v)=rV,\qquad W(v)=\omega V.
\end{equation}
Now Eq.~\eqref{eq:observable-main} becomes
\begin{equation}\label{eq:mode-substitution}
\dot{\sigma}\geq
\frac{2\omega V}{\rho(v)^2\sin(2\pi/d)}
\artanh\!\left(\mathcal R\tan\frac\pi d\right).
\end{equation}
Multiplying by $2\pi/\omega$ and setting
$\etad=V/\rho(v)^2$ gives Eq.~\eqref{eq:spectral-main}.
The range $\mathcal R<\cot(\pi/d)$ follows directly from Eq.~\eqref{eq:cycle-domain} through the observable theorem.

For completeness, the geometric factor has the stated bounds and equality condition.
Because $c=0$ is an allowed enclosing-disk center,
\begin{equation}\label{eq:eta-compare}
\rho(v)^2\leq\max_i|v_i|^2,
\qquad
\eta=\frac{V}{\max_i|v_i|^2}\leq\etad.
\end{equation}
Conversely, use $\sum_ip_i^{\rm ss} v_i=0$ to expand the weighted squared distance from any center $c$:
\begin{equation}\label{eq:mean-disk}\begin{aligned}
\sum_ip_i^{\rm ss}|v_i-c|^2
&=V+|c|^2\\
&\leq\max_i|v_i-c|^2.
\end{aligned}\end{equation}
Choose a center $c_*$ minimizing the maximum distance.
Equation~\eqref{eq:mean-disk} then yields
\begin{equation}\label{eq:eta-upper}
V+|c_*|^2\leq\rho(v)^2,
\qquad 0<\etad\leq1.
\end{equation}
This also explains why a substantial fraction of near-zero mode amplitudes can reduce $\etad$ even when all states have equal stationary probabilities.

If $\etad=1$, then $\rho(v)^2=V$ forces $c_*=0$ and equality between the weighted average $\sum_ip_i^{\rm ss}|v_i|^2$ and its maximum.
Since every $p_i^{\rm ss}>0$, all $|v_i|^2$ must equal $V$.
Conversely, if all $|v_i|$ are constant, Eq.~\eqref{eq:mean-disk} and the admissible center $c=0$ give
$V\leq\rho(v)^2\leq\max_i|v_i|^2=V$.
Thus $\etad=1$ if and only if $v$ has constant modulus.

Finally, for the Fourier observable on the uniform ring of Eq.~\eqref{eq:ring-construction}, direct application of the transition rate matrix gives
\begin{equation}\label{eq:ring-eigenvalues}\begin{aligned}
r&=(k_++k_-)(1-\cos\theta),\\
\omega&=(k_+-k_-)\sin\theta.
\end{aligned}\end{equation}
Here $V=\rho^2$ and $\etad=1$, so the same ring also saturates the spectral inequality at every admissible $\mathcal R$.
This establishes sharpness of the full homogeneous-ring curve for constant-modulus modes, without claiming sharpness for an arbitrarily prescribed factor $\etad<1$.

\section{Finite-time correlations and endpoint entropy production}\label{app:finite-time}

Here we derive Eq.~\eqref{eq:finite-time-main} for the same stationary
finite-state Markov jump process and observable pair used in the main
text.  Let
\begin{equation}\label{eq:finite-time-matrix}
P(\tau)=e^{R\tau},\qquad
R^{(\tau)}=\frac{P(\tau)-I}{\tau},\qquad \tau>0.
\end{equation}
Our matrix convention is that $P_{ji}(\tau)$ is the probability of
finding the system in state $j$ at time $\tau$ when it started in
state $i$.  For $i\ne j$,
$R^{(\tau)}_{ji}=P_{ji}(\tau)/\tau\geq0$, and
$\sum_jR^{(\tau)}_{ji}=0$.  The original stationary distribution is
preserved because $P(\tau)\boldsymbol p^{\rm ss}
=\boldsymbol p^{\rm ss}$.
Irreducibility implies $P_{ji}(\tau)>0$ for every pair of states and
$\tau>0$. The auxiliary matrix therefore also has positive rates in
both directions.  Thus $R^{(\tau)}$ satisfies the assumptions used in
the proof of Eq.~\eqref{eq:observable-main}.  It is an auxiliary
transition rate matrix, not a statement that the microscopic system
makes at most one jump during a sampling interval.

Write $C^{(\tau)}_{fg}(t)$ for the stationary correlation of the
auxiliary process, to distinguish its time variable $t$ from the
observation lag $\tau$ of the original process.  By the short-time
identity of Eq.~\eqref{eq:correlation-derivative},
\begin{equation}\label{eq:finite-time-slopes}
\begin{aligned}
\left.\frac{d}{dt}C^{(\tau)}_{fg}(t)\right|_{t=0^+}
&=\sum_i p_i^{\rm ss}f_i\bigl[(R^{(\tau)})^{\mathsf T}g\bigr]_i\\
&=\frac{C_{fg}(\tau)-C_{fg}(0)}{\tau}.
\end{aligned}
\end{equation}
In particular, since $C_{ab}(0)=C_{ba}(0)$,
\begin{equation}\label{eq:finite-time-effective-DW}
\begin{aligned}
D_a^{(\tau)}&=\frac{\Delta_a(\tau)}{\tau},\qquad
D_b^{(\tau)}=\frac{\Delta_b(\tau)}{\tau},\\
2\alpha_{ab}^{(\tau)}&=\frac{\chi_{ab}(\tau)}{\tau}.
\end{aligned}
\end{equation}
The observables themselves are unchanged, so the support count $d$
and enclosing-disk radius $\rho$ of the auxiliary process are the
same as for the original process.  Applying
Eq.~\eqref{eq:observable-main} to $R^{(\tau)}$ and using
Eq.~\eqref{eq:finite-time-effective-DW} yields
\begin{equation}\label{eq:finite-time-aux-bound}
\dot{\sigma}^{(\tau)}\geq
\frac{2|\chi_{ab}(\tau)|}{\tau\rho^2\sin(2\pi/d)}
\artanh\!\left(
\frac{|\chi_{ab}(\tau)|}{\Delta_a(\tau)+\Delta_b(\tau)}
\tan\frac{\pi}{d}\right).
\end{equation}
The strict domain for the argument of $\artanh$ follows from the
observable-value constraint of Eq.~\eqref{eq:observable-domain}
applied to the auxiliary process.  The denominator in
Eq.~\eqref{eq:finite-time-aux-bound} is positive whenever the
observable pair is nonconstant: for $\tau>0$, the positive transition
probabilities make the autocorrelation losses positive.

It remains to compare $\dot{\sigma}^{(\tau)}$ with the actual entropy
production $\dot{\sigma}$.  Define the steady-state probability of an
observed initial--final state pair by
\begin{equation}\label{eq:finite-time-pairs}
\nu_{ij}(\tau)=p_i^{\rm ss}P_{ji}(\tau).
\end{equation}
The forward and reverse steady-state fluxes of the auxiliary process
are $\nu_{ij}(\tau)/\tau$ and $\nu_{ji}(\tau)/\tau$.
Consequently,
\begin{equation}\label{eq:finite-time-aux-sigma}
\begin{aligned}
\tau\dot{\sigma}^{(\tau)}
&=\sum_{i<j}[\nu_{ij}(\tau)-\nu_{ji}(\tau)]
\ln\frac{\nu_{ij}(\tau)}{\nu_{ji}(\tau)}\\
&=D_{\mathrm{KL}}\!\left(\nu(\tau)\middle\|
\nu^{\mathsf T}(\tau)\right).
\end{aligned}
\end{equation}
Here $D_{\mathrm{KL}}$ is the relative entropy between the two
normalized distributions of ordered endpoint pairs. The equal-state pairs contribute zero.

For a stationary Markov jump process, the relative entropy between the
distribution of complete trajectories on $[0,\tau]$ and that of their
time reversals is $\tau\dot{\sigma}$.
Retaining only the initial and final states maps these trajectory
distributions to $\nu(\tau)$ and its reversal
$\nu^{\mathsf T}(\tau)$.  The log-sum inequality (equivalently,
relative-entropy contraction under discarding trajectory information)
therefore gives
\begin{equation}\label{eq:finite-time-contraction}
\tau\dot{\sigma}\geq
D_{\mathrm{KL}}\!\left(\nu(\tau)\middle\|
\nu^{\mathsf T}(\tau)\right)
=\tau\dot{\sigma}^{(\tau)}.
\end{equation}
Combining Eqs.~\eqref{eq:finite-time-aux-bound} and
\eqref{eq:finite-time-contraction} proves
Eq.~\eqref{eq:finite-time-main}.  This also makes clear why equality
at nonzero lag is not guaranteed: endpoint statistics generally omit
some of the irreversibility of trajectories with intermediate jumps.
The argument assumes stationarity throughout. It does not establish
the same bound for general time-dependent driving.

\bibliography{references}
\end{document}